\documentstyle[12pt]{article}
\begin{document}
 \bibliographystyle{unsrt}
 \vbox{\vspace{6mm}}

\begin{center}
{\Large {\bf Symplectic tomography of nonlinear coherent states
of a trapped ion}}
\end{center}
\bigskip

\begin{center}
{\bf O.V. Man'ko}
\end{center}
\medskip

\begin{center}
{\it Lebedev Physical Institute, Leninsky Pr., 53, 117924 Moscow, Russia}
\end{center}
\bigskip

\begin{abstract}

\noindent

Squeezed and rotated quadrature of an ion in a Paul trap is
discussed in connection with reconstructing its quantum state
using symplectic-tomography method. Marginal distributions
of the quadrature for squeezed and correlated states and for
nonlinear coherent states of a trapped ion are obtained and the
density matrices in the Fock basis are expressed explicitly in
terms of these marginal distributions.
\end{abstract}

\section{Introduction}

\noindent

Recently,  such nonclassical states of a trapped ion as even and
odd coherent states~\cite{Physica74} (or Schr\"odinger cat
states~\cite{Knight}\,) were realized experimentally~\cite{Wineland}
 and their properties were discussed in Refs.~[4--6].
For light modes, these states were produced in high-$Q$
cavities~\cite{Haroche}. The  experiments with reproducible measurements
of squeezed vacuum state of light generated by an optical parametric
oscillator were performed in Ref.~\cite{Mlynek}. Resonance fluorescence
was proposed to reconstruct the quantum-mechanical state of a trapped
ion~\cite{W.V.}.

Endoscopy method of measuring the nonclassical states, in particular,
of a trapped ion was suggested in Ref.~\cite{Schleich} and tomography
method for studying the Schr\"odinger cat states of an ion in a Paul
trap was discussed in Ref.~\cite{Olga1}. In Refs.~\cite{Glaconf,Octavio},
the linear integrals of motion of the parametric
oscillator~\cite{Vol183,MM70}, which models the motion of a trapped
ion, were used to study the quantum states of the system.

New type of nonclassical states, namely, nonlinear coherent states of an
ion and the method of creation of these states, as stationary states
of the center-of-mass motion of a trapped and bichromatically
laser-driven ion, were suggested in Ref.~\cite{Vogelnl}. The nonlinear
states are the particular case of f-coherent states introduced in
Ref.~\cite{Sudarshan} to describe a nonlinear quantum oscillator, for
which the phase of vibrations depends on the energy of the vibrations.
In the linear limit, these nonclassical states, which have the properties 
of squeezing and correlation of quadratures, become the
coherent states of harmonic oscillator~\cite{Glauber63}. Thus, the
problem of experimental reconstructing the nonclassical state in terms
of Wigner function or density matrix (in other representations) for the
nonlinear coherent state of a trapped ion is actual problem.

Recently, the symplectic tomography method was discussed for measuring
a quantum state~\cite{Mancini1}. The method (extended as well for
multimode case~\cite{Ariano}\,) uses Fourier transform of marginal
distribution for measurable squeezed and rotated quadrature instead
of Radon transform~\cite{VogRis}, which is used in optical tomography
to reconstruct the Wigner function; in this context, the symplectic
tomography is similar to the strength-field method of Ref.~\cite{Vogel94}.
The marginal distribution for squeezed and rotated quadrature determines
completely the quantum state and measuring this distribution implies
reconstructing the quantum state. It satisfies the classical-like
evolution equation introduced in Ref.~\cite{Mancini2} (see, also
review~\cite{MRLR}\,).

The aim of this work is to discuss the symplectic tomography scheme to
measure nonlinear coherent states of a trapped ion, following the approach
considered in Ref.~\cite{Olga1} for measuring even and odd coherent
states. Our goal is to construct explicitly the marginal distribution of
squeezed and rotated quadrature for nonlinear coherent state of an ion in
a Paul trap and to compare it with marginal distribution of optical
tomography procedure. As well, the marginal distribution for discrete
oscillator levels will be considered, corresponding to photon-number
tomography of Refs.~[24--26],
as a procedure for measuring the quantum state of the trapped ion.

\section{Nonlinear coherent states of a trapped ion}

\noindent

Since an ion in a Paul trap is described by the model of a parametric
oscillator~\cite{Glaconf,SchraMa} in this section we review its properties.
For a parametric oscillator with an arbitrary time dependence of the
frequency and the Hamiltonian
\begin{equation}\label{ss1}
H=-\frac {\partial ^{2}}{2\,\partial x^{2}}
+\frac {\omega ^{2}\,(t)\,x^{2}}{2}\,,
\end{equation}
where we put $\hbar=m=\omega \,(0)=1$ and used expressions for the position
and momentum operators in the coordinate representation, there is the
time-dependent integral of motion found in Ref.~\cite{MM70}\,:
\begin{equation}\label{ss2}
A=\frac {i}{\sqrt 2}\,\left [\,\varepsilon \,(t)\,\hat p
-\dot \varepsilon \,(t)\,\hat q\,\right ],
\end{equation}
where
\begin{equation}\label{ss3}
\ddot \varepsilon \,(t)+\omega ^{2}\,(t)\,\varepsilon \,(t)=0\,;
\qquad \varepsilon \,(0)=1\,;\qquad \dot \varepsilon \,(0)=i\,,
\end{equation}
which satisfies the commutation relation
\begin{equation}\label{ss4}
[A,\,A^\dagger ]=1\,.
\end{equation}
For the trapped ion, the time dependence of the frequency is taken to be
periodic~\cite{Glaconf}\,:
\begin{equation}\label{insert1}
\omega ^2 \,(t)=1+\kappa ^2\sin ^2\Omega t\,.
\end{equation}
It is easy to show that Gaussian packet solutions to the Schr\"odinger
equation may be introduced and interpreted as coherent states~\cite{MM70},
since they are eigenstates of the operator $A$~(\ref{ss2}), of the form
\begin{equation}\label{ss5}
\Psi _\alpha \left (x,\,t\right )=\Psi _0\left (x,\,t\right )\exp
\left \{-\frac {|\alpha |^{2}}{2}-
\frac {\alpha ^{2}\varepsilon ^{*}(t)}{2\varepsilon (t)}
+\frac {{\sqrt 2}\alpha x}{\varepsilon}\right \},
\end{equation}
where
\begin{equation}\label{ss6}
\Psi _0\left (x,\,t\right )=\pi ^{-1/4}\left [\varepsilon (t)
\right ]^{-1/2}\exp \frac {i\dot \varepsilon (t)x^{2}}
{2\varepsilon (t)}
\end{equation}
is an analog of the ground state of the oscillator and $\alpha $ is a
complex number. The variances of the position and momentum of the
parametric oscillator in the state~(\ref{ss5}) are
\begin{equation}\label{ss7}
\sigma _{qq}=\frac {|\varepsilon \,(t)|^{2}}{2}\,;\qquad
\sigma _{pp}=\frac {|\dot \varepsilon \,(t)|^{2}}{2}\,,
\end{equation}
and the correlation coefficient $r$ of the position and momentum has
a value corresponding to minimization of the Schr\"odinger uncertainty
relation~\cite{Schr30}\,:
\begin{equation}\label{ss8}
\sigma _{qq}\,\sigma _{pp}=\frac {1}{4}\,\frac {1}{1-r^{2}}\,;
\qquad
r=\frac {\sigma _{pq}}{\sqrt {\sigma _{qq}\,\sigma _{pp}}}\,.
\end{equation}
If $\sigma _{qq}<1/2~~(\sigma _{pp}<1/2)\,,$ we have squeezing in
quadrature components.

Analogs of an orthogonal and complete system of number states, which
are excited states of an ion in a Paul trap, are obtained by expansion
of~(\ref{ss5}) into a power series in $\alpha .$ We have
\begin{equation}\label{ss9}
\Psi _m\left (x,\,t\right )=\left (\frac {\varepsilon ^*(t)}
{2\varepsilon (t)}\right )^{m/2}\frac {1}{\sqrt {m!}}\,
\Psi _0\left (x,\,t\right )H_m\left (
\frac {x}{|\varepsilon (t)|}\right ),
\end{equation}
and these squeezed and correlated number states are eigenstates of the
invariant $A^{\dagger }A.$

The coherent state~(\ref{ss5}), which is squeezed and correlated state
for quadratures, is the superposition of number states~(\ref{ss9})
\begin{equation}\label{ola1}
\Psi _\alpha \left (x,\,t\right )=\exp \left (-\frac {|\alpha |^2}{2}
\right )\sum _{m=0}^\infty \frac {\alpha ^m}{\sqrt {m!}}\,
\Psi _m\left (x,\,t\right ).
\end{equation}
There exist the integrals of motion
\begin{equation}\label{AR7}
B=A\,f\left (A^\dagger A\right );\qquad
B^\dagger =f\left (A^\dagger A\right )A^\dagger \,,
\end{equation}
which are determined by a function $f$ of the invariants~(\ref{ss2}).
These integrals of motion satisfy the commutation relations
\begin{equation}\label{ola2}
\left [B,B^\dagger \right ]=F\left (A^\dagger A\right ),
\end{equation}
where
\begin{equation}\label{AR11}
F\left (A^\dagger A\right )=\left (A^\dagger A+1\right )f^2\left (
A^\dagger A+1\right )-A^\dagger A\,f^2\left (A^\dagger A\right ).
\end{equation}
Generalizing notion of coherent states to the case of the operator,
which is nonlinearly transformed annihilation operator, we introduce
the eigenfunctions of the invariant $B$
\begin{equation}\label{ola3}
B\,\Psi _\beta \left (x,\,t\right )
=\beta \,\Psi _\beta \left (x,\,t\right ),
\end{equation}
which are the nonlinear coherent states. Such construction of the
states, called f-coherent states, was suggested in Ref.~\cite{Sudarshan}.
For the function
\begin{equation}\label{ola4}
f(y)=L_{y+1}^1(\eta ^2)\left [yL_{y+1}^0(\eta ^2)\right ]^{-1},
\end{equation}
where $L_m^n(\eta ^2)$ are assosiated Laguerre polynomials and $\eta~$ is
Lamb--Dicke parameter, the f-coherent states (nonlinear
coherent states) have been considered in Ref.~\cite{Vogelnl}.

Using general scheme of constructing the normalized nonlinear coherent
states of Refs.~\cite{Vogelnl,Sudarshan} one can obtain the function
$\Psi _\beta (x,t)$ in the form of series
\begin{equation}\label{ola5}
\Psi _\beta \left (x,\,t\right )=\left (\sum _{n=0}^\infty
\frac {|\beta |^{2n}}{n!\,|[f(n)]!|^2}\right )^{-1/2}
\sum _{m=0}^\infty \frac {\beta ^m}{\sqrt {m!}\,[f(m)]!}\,
\Psi _m\left (x,\,t\right ),
\end{equation}
in which we denote, e.g., $[f(m)]!=f(0)f(1)\cdots f(m)\,.$

For $f(m)=1\,,$ the wave function of the nonlinear coherent
state~(\ref{ola5}) becomes the wave
function of the coherent state~(\ref{ss5}), in which the parameter
$\beta =\alpha \,.$ The Wigner function of the nonlinear coherent
states~(\ref{ola5}) has the form~\cite{Sudarshan}
\begin{eqnarray}\label{FQ27}
W_\beta \left (x,\,p\right )&=&2\left (\sum _{n=0}^\infty
\frac {|\beta |^{2n}}{n!\,|[f(n)]!|^2}\right )^{-1}
e^{-(x^2 + p^2)}\nonumber\\
&&\times \sum _{m=0}^\infty \sum _{n=0}^\infty
\frac {1}{m!\,[f(m)]!}\,\frac {1}{[f(n)]!}\,
(-\beta )^n\beta ^{*m}\nonumber\\
&&\times \left (\sqrt 2\,[x-ip]\right )^{m-n}\,
L_n^{m-n}\left (2\left [x^2+p^2\right ]\right ),
\end{eqnarray}
where $L_m^n$ denotes associated Laguerre polynomial. For the particular
case of the function $f$ given by~(\ref{ola4}), one has the Wigner
function studied in Ref.~\cite{Vogelnl}, where some plots of the
Wigner function were presented.

\section{Tomography of a trapped ion}

\noindent

It was shown~\cite{Mancini1} that for the  generic linear combination
of quadratures, which is a measurable observable
$\left (\hbar =1\right),$
\begin{equation}\label{X}
\widehat X=\mu \hat q+\nu\hat p\,,
\end{equation}
where $\hat q$ and $\hat p$ are the position and momentum,
respectively, the marginal distribution
$w\,(X,\,\mu,\,\nu )$ (normalized with respect to the
variable $X$), depending on the two extra real parameters
$\mu $ and $\nu \,,$ is related to the state of the quantum system
expressed in terms of its Wigner function $W(q,\,p)$ as follows
\begin{equation}\label{w}
w\left (X,\,\mu,\,\nu \right )=\int \exp \left [-ik(X-\mu q-\nu
p)\right ]W(q,\,p)\,\frac {dk\,dq\,dp}{(2\pi)^2}\,.
\end{equation}
The physical meaning of the parameters $\mu $ and $\nu $ is that
they describe an ensemble of rotated and scaled reference frames,
in which the position $X$ is measured. For $\mu =\cos \,\varphi ;
\,\nu =\sin \,\varphi ,$ the marginal distribution~(\ref{w}) is
the distribution for homodyne output variable used in optical
tomography~\cite{VogRis}. Formula~(\ref{w}) can be inverted and
the Wigner function of the state can be expressed in terms of the
marginal distribution~\cite{Mancini1}\,:
\begin{equation}\label{W}
W(q,\,p)=\frac {1}{2\pi }\int w\left (X,\,\mu ,\,\nu \right )
\exp \left [-i\left (\mu q+\nu p-X\right )\right ]
\,d\mu \,d\nu \,dX\,.
\end{equation}
It was shown~\cite{Mancini2} that for systems with the Hamiltonian
of the form $\hat H=(\hat p^2/2)+V(\hat q)$ the marginal distribution
satisfies quantum time-evolution equation. For a trapped ion,
the evolution equation takes the form
\begin{equation}\label{TIE}
\dot w-\mu \,\frac {\partial}{\partial \nu }\,w+\omega ^2\left (t\right )
\nu \,\frac {\partial }{\partial \mu }\,w=0\,.
\end{equation}
If one uses the constraint $\mu =\cos \,\varphi ;\,\nu =\sin
\,\varphi ,$ this equation becomes the equation for marginal
distribution of optical tomography~\cite{VogRis}. Thus, measuring
marginal distribution for scaled and rotated quadrature one can
reconstruct the Wigner function of a trapped ion using the Fourier
transform~(\ref{W}).

\section{Marginal distribution for squeezed and correlated states}

\noindent

First we discuss marginal distribution for squeezed and correlated
state~(\ref{ss5}) of a trapped ion. For these states, the Wigner
function has Gaussian form~\cite{Olga2}. Consequently, the Fourier
transform~(\ref{w}) of the Gaussian Wigner function determining the
marginal distribution yields the Gaussian form of this marginal
distribution
\begin{equation}\label{freesolution}
w_\alpha \left (X,\,\mu ,\,\nu ,\,t\right )=
\frac{1}{\sqrt{2\pi \sigma_X(t)}}\,\exp \left\{-\frac{(X-\bar {X})^2}
{2\sigma _X(t)}\right\},
\end{equation}
in which, in view of (\ref{X}) and (\ref{ss2}), one has
\begin{equation}\label{insert7}
\bar {X}=\mu \langle q\rangle +\nu \langle p\rangle \,,
\end{equation}
where quadrature means are
\begin{eqnarray}
\langle p\rangle &=&\frac {1}{\sqrt 2}
\left (\alpha \dot \varepsilon ^*+\alpha ^*\dot \varepsilon \right)\,;
\label{insert5a}\\
\langle q\rangle &=&\frac {1}{\sqrt 2}\left (\alpha \varepsilon ^*
+\alpha ^*\varepsilon \right)\,.\label{insert5b}
\end{eqnarray}
The dispersion and correlation of the quadratures are
\begin{equation}\label{insert6}
\sigma _X(t)=\mu ^2\sigma _{qq}+\nu ^2\sigma _{pp}
+2\mu \nu \sigma _{pq}\,,
\end{equation}
where the parameters $\sigma _{qq}\,,\,\sigma _{pp}\,,$ and
$\sigma _{pq}$ are given by~(\ref{ss7}) and (\ref{ss8}).
It is easy to check that the marginal distribution for the
nonclassical state~(\ref{freesolution}) satisfies a
classical-like evolution equation for the density matrix (\ref{TIE})
introduced in symplectic tomography scheme.

\section{Marginal distribution for nonlinear coherent states}

\noindent

Using the decomposition of wave function of nonlinear coherent
state into series~(\ref{ola5}) one can obtain by standard method
the Wigner function of the trapped ion in the form
\begin{eqnarray}\label{FQ27b}
W_\beta \left (q,\,p,\,t\right )&=&2\left (\sum _{n=0}^\infty
\frac {|\beta |^{2n}}{[n]!}\right )^{-1}\,\exp
\left (-\left [q^2(t)+p^2(t)\right ]\right )\nonumber\\
&&\times \sum _{m=0}^\infty \sum _{n=0}^\infty
\frac {1}{m!\,[f(m)]!}\,\frac {1}{[f(n)]!}
(-\beta )^n\beta ^{*m}\nonumber\\
&&\times \left (\sqrt {2}\left [q(t)-ip(t)\right ]\right )^{m-n}
\,L_n^{m-n}\left (2\left [q^2(t)+p^2(t)\right ]\right ),
\end{eqnarray}
where
\begin{eqnarray}
p\,(t)&=&\frac {\varepsilon +\varepsilon ^*}{2}\,p-
\frac {\dot \varepsilon +\dot \varepsilon ^*}{2}\,q\,;\label{xxx1}\\
q\,(t)&=&-\frac {\varepsilon -\varepsilon ^*}{2i}\,p+
\frac {\dot \varepsilon -\dot \varepsilon ^*}{2i}\,q\,.\label{xxx2}
\end{eqnarray}
Then calculating the Fourier integral~(\ref{w}) one obtains the
marginal distribution of the trapped ion in nonlinear coherent state
in the form
\begin{equation}\label{ola6}
w_\beta \left (X,\,\mu ,\,\nu ,\,t\right )=\left (\sum _{n=0}^\infty
\frac {|\beta |^{2n}}{[n]!}\right )^{-1}\,\sum _{m=0}^\infty
\sum _{n=0}^\infty \frac {\beta ^n\beta ^{*m}}{\sqrt {n!}\,\sqrt {m!}
\,[f(n)]!\,[f(m)]!}\,w_{nm}\left (X,\,\mu ,\,\nu ,\,t\right ).
\end{equation}
Here $w_{nm}\left (X,\,\mu ,\,\nu ,\,t\right )$ is
\begin{eqnarray}\label{eso12}
w_{nm}\left (X,\,\mu ,\,\nu ,\,t\right )&=&\left (\pi \left [
\mu ^2(t)+\nu ^2(t)\right ]\right )^{-1/2}
\,2^{-(n/2+m/2)}\,(n!\,m!)^{-1/2}\nonumber\\
&&\times \frac {\left [\nu (t)+i\mu (t)\right ]^n\,
\left [\nu (t)-i\mu (t)\right ]^n}{\left [\mu ^2(t)
+\nu ^2(t)\right ]^{n/2+m/2}}
\exp \,\left (-\frac {X^2}{\mu ^2(t)+\nu ^2(t)}\right )\nonumber\\
&&\times
H_n\left (\frac {X}{\sqrt {\mu ^2(t)+\nu ^2(t)}}\right )
H_m\left (\frac {X}{\sqrt {\mu ^2(t)+\nu ^2(t)}}\right ),
\end{eqnarray}
where
\begin{eqnarray}
\nu \,(t)&=&\frac {\dot \varepsilon -\dot \varepsilon ^*}{2i}\,\nu
+\frac {\varepsilon -\varepsilon ^*}{2i}\,\mu \,;\label{xxxx1}\\
\mu \,(t)&=&\frac {\dot \varepsilon +\dot \varepsilon ^*}{2}\,\nu
+\frac {\varepsilon +\varepsilon ^*}{2}\,\mu \,.\label{xxxx2}
\end{eqnarray}
One can check that the marginal distribution function~(\ref{ola6})
is the solution to quantum evolution equation~(\ref{TIE}).
For $f\left (n\right )=1$ and $\beta =\alpha \,,$ the marginal
distribution function~(\ref{ola6}) coincides with the Gaussian
distribution~(\ref{freesolution}).

The method of photon number tomography was discussed
in Ref.~\cite{Mancini3}. In this method, the measurable marginal
distribution $w\left (n,\,\alpha \right )$ depends on the discrete
number of photons $n$ and on the complex amplitude $\alpha $ of the
local field oscillator, that may be scanned. Using the general
relation of symplectic marginal distribution to the distribution
of the discrete photon number~\cite{Jmodopt}, one can obtain
the discrete marginal distribution
$w_\beta \left (n,\,\alpha ,\,t\right )$ for an ion in a Paul trap, which
is modeled by a parametric oscillator, in terms of the marginal
distribution~(\ref{w}) for the nonlinear coherent states in the form
\begin{eqnarray}\label{C}
w_\beta \left (n,\,\alpha ,\,t\right )&=&\frac {1}{2\pi }\int
w_\beta \left (X,\,\mu ,\,\nu ,\,t\right )
\exp \left \{iX-\frac {\mu ^2 +\nu ^2}{4}
+\frac {\alpha (\nu +i\mu )}{\sqrt 2}
-\frac {\alpha ^* (\nu -i\mu )}{\sqrt 2}\right \}\nonumber\\
&&\times L_n\left (\frac {\mu ^2+\nu ^2}{2}\right )
\,dx\,d\mu \,d\nu \,,
\end{eqnarray}
where $w_\beta \left (X,\,\mu ,\,\nu ,\,t\right )$
is given by~(\ref{ola6}).

The endoscopy method~\cite{Schleich} was used to reconstruct
matrix elements of density operator of an ion in a Paul trap
$\langle m|\rho |n\rangle $ in the Fock basis on the example
of damped odd coherent state. We show how these matrix elements
may be reconstructed following the approach of the symplectic
tomography method~\cite{Mancini1,Ariano,Jmodopt}. It was
demonstrated~\cite{Ariano} that the density operator is expressed
as convolution of the marginal distribution and displacement
operator creating the coherent state from the vacuum state.
Since the matrix elements of the displacement operator in the
Fock basis are well known, the formula for the matrix elements
of the density operator in the Fock basis may be expressed in
terms of assosiated Laguerre polynomials~\cite{Jmodopt}.
Thus, for nonlinear coherent state of an ion in a Paul trap
one has the expression for the density matrix in the Fock basis
as convolution of the marginal distribution for squeezed and
rotated quadrature and the assosiated Laguerre polynomial
\begin{eqnarray}\label{ncsti}
\langle m|\rho_\beta \left (t\right )|n\rangle &=&\sqrt {\frac {n!}{m!}}
\,\frac {2^{(n-m)/2}}{2\pi }\int \exp \left (ix-\frac {\mu ^2
+\nu ^2}{4}\right )
w_\beta \left (X,\,\mu ,\,\nu ,\,t\right )
\left (\nu -i\mu \right )^{m-n}\nonumber\\
&&\times L_n^{m-n}\left (\frac {\nu ^2
+\mu ^2}{2}\right )dX\,d\mu \,d\nu \,;\qquad m>n\,.
\end{eqnarray}
Here the function $w_\beta \left (X,\,\mu ,\,\nu ,\,t\right )$ is
given by~(\ref{ola6}). For $m<n\,,$ we use the property
of the hermitian density matrix $\langle m|\rho_\beta \left (
t \right )|n\rangle =\langle n|\rho_\beta \left (t\right )|m\rangle ^*.$

\section{Conclusion}

\noindent

One can conclude that for an ion in a Paul trap there exist three types
of different marginal distributions, which can be measured
experimentally. The explicit expressions for the distributions for squeezed
states and for nonlinear coherent states are the main result
of our work. These marginal distributions are

(i)~~the distribution for squeezed and rotated quadrature;

(ii)~~the distribution for homodyne observable, which is partial
case of the squeezed and rotated quadrature for the parameters
$\mu =\cos \,\varphi ,\,\nu =\sin \,\varphi \,;$

(iii)~~the number distribution controlled by the scanned complex
amplitude of vibrations of a local oscillator.

Measurement of any of the three distributions gives the
reconstruction of the quantum state of the trapped ion.
The explicit expressions for the
marginal distributions obtained in this work yield the theoretical
predictions for measuring experimentally the nonlinear coherent states
of an ion in a Paul trap with the help of the three different methods,
namely, optical tomography, symplectic tomography, and photon number
tomography.

\section*{Acknowledgments}

\noindent

This work has been partially supported by the Russian Foundation for
Basic Research under the Project No.~96-02-17222. The author thanks
Prof.~H.~Paul, Prof.~A.~W\"unsche, and Prof.~V.I.~Man'ko for fruitfull
discussions.

\end{document}